\documentclass[lnbip]{svmultln}
\usepackage{graphicx}
\usepackage{makeidx}  % allows for indexgeneration
\usepackage{multirow}
\usepackage{float}
\begin{document}
\mainmatter              % start of the contribution
\title{Moving Beyond the Mean: Analyzing Variance in Software Engineering Experiments}
\titlerunning{Moving Beyond the Mean}  % abbreviated title (for running head)
\author{Adrian Santos\inst{1} \and Markku Oivo\inst{1} \and Natalia Juristo\inst{2}}
\authorrunning{Santos et al.}   % abbreviated author list (for running head)
\tocauthor{Adrian Santos, Markku Oivo, Natalia Juristo}
\institute{M3S (M-Group), ITEE University of Oulu, Finland,\\
\email{adrian.santos.parrilla/markku.oivo@oulu.fi},
\and
Escuela T\'ecnica Superior de Ingenieros Inform\'aticos, Universidad Polit\'ecnica de Madrid, Spain,\\
\email{natalia@fi.upm.es}}

\maketitle              % typeset the title of the contribution
% \index{Ekeland, Ivar} % entries for the author index
% \index{Temam, Roger}  % of the whole volume
% \index{Dean, Jeffrey}

\begin{abstract} 
% give a summary of your paper in 150 words
Software Engineering (SE) experiments are traditionally analyzed with statistical tests (e.g., $t$-tests, ANOVAs, etc.) that assume equally spread data across groups (i.e., the homogeneity of variances assumption). Differences across groups' variances in SE are not seen as an opportunity to gain insights on technology performance, but instead, as a hindrance to analyze the data. We have studied the role of variance in mature experimental disciplines such as medicine. We illustrate the extent to which variance may inform on technology performance by means of simulation. We analyze a real-life industrial experiment on Test-Driven Development (TDD) where variance may impact technology desirability. Evaluating the performance of technologies just based on means---as traditionally done in SE---may be misleading. Technologies that make developers obtain similar performance (i.e., technologies with smaller variances) may be more suitable if the aim is minimizing the risk of adopting them in real practice.

% please supply keywords within your abstract
\end{abstract}
\section{Introduction}
SE experiments are traditionally analyzed with statistical tests (e.g., $t$-tests, ANOVAs, etc. \cite{dybaa2006systematic}) that assume equally spread data across groups (i.e., the homogeneity of variances assumption \cite{field2013discovering}). Perhaps inadvertently, and as a consequence of just relying on traditional statistical tests' results, researchers judge the performance of software technologies solely with regard to their mean performances. Not just differences across mean performances are relevant when deciding on the suitability of a new technology. Variation of technologies' scores, for example, may also be relevant if the aim is minimizing the risk of adopting such technologies in real-life contexts.

For example, let us suppose that two development processes (let us say Method A and Method B) perform similarly 'on-average' (and thus, that the estimation of the \textit{means} of Method A and Method B are similar) on a certain outcome of interest (e.g., quality in a percentage scale) in an experiment where two independent groups of developers apply each a different development process (i.e., an AB between-subjects experiment \cite{wohlin2012experimentation, juristo2013basics}). Let us further suppose that albeit both groups achieve similar means, the spread of the scores in each group are different (e.g., Method A's quality scores are clumped together close to the mean, and Method B's quality scores are largely dispersed along the 0-100\% interval). Even though Method A and Method B perform similarly 'on-average' (i.e., in terms of means), developers' quality scores with Method B are more spread than those with Method A. A traditional statistical test (e.g., a $t$-test \cite{field2013discovering}) applied on such data will not detect any difference between the means of both methods (as after all, 'on-average' both methods perform similarly, and $t$-tests just compare means \cite{field2013discovering}). Does this imply that Method A and Method B are equally suitable in all circumstances? If we had to make a decision and choose a technology for a group of developers, which one would we prefer?

It depends. If we were a risk-averse manager, choosing Method A (i.e., the less variable method) over Method B may be beneficial. After all, as all developers are expected to obtain similar quality scores with Method A, it is possible to make precise predictions on the quality to be achieved in a new software product being developed. On the contrary, if we were an enthusiastic developer, choosing Method B (i.e., the most variable method) may be beneficial. Indeed, it might be the case that we obtain large benefits with it, certainly not a thing to expect with Method A---as all developers obtain quality scores close to the mean.

Along this work we aim to answer a main \textbf{research question}:
\begin{itemize}
    \item{Is it worth it investigating variance in SE experiments?} 
\end{itemize}

To answer this research question we first perform a simulation to illustrate the extent to which different variances across technologies---even if their means are identical---may determine their suitability. Then, we analyze a real-life industrial experiment on TDD where a similar circumstance may have materialized. In particular, by analyzing the data just with traditional statistical tests (i.e., the $t$-test) as it is commonly done in SE experiments, both technologies seem to perform similarly. To what extent is it so? Along this research we found:
\\
\\
\noindent\fbox{
  \parbox{\columnwidth}{
    \textbf{Key findings}
    \begin{itemize}
        \item{Not just differences between means are relevant when judging technology's performance: differences between variances may be relevant also.}
        \item{Failing to detect real differences across treatments' variances does not mean that technologies have identical variances: statistical tests may be under-powered to detect real differences across variances given SE experiments' small sample sizes.}
    \end{itemize}  
  }
}
\\\\\\
The main \textbf{contributions} of this paper are a \textit{a call to analyze variance in SE experiments} and a reminder \textit{that other statistical points rather than means (e.g., variances) may also serve to inform about technology performance}. 

Along this study we argue that as SE experiments are commonly analyzed by means of traditional statistical tests, perfectly suitable technologies may have passed unnoticed due to their perhaps similar mean performances. In addition, as individual experiments in SE are generally small to detect real differences between means \cite{dybaa2006systematic}, the same may hold for detecting differences between variances \cite{quinn2002experimental}. In view of this, we suggest:
\\
\\
\noindent\fbox{
  \parbox{\columnwidth}{
    \textbf{Actionable results}
    \begin{itemize}
        \item{Variance should be investigated in SE experiments and considered when judging technology performance.}
        \item{Effect sizes quantifying the difference between technologies' variances should be provided with the aim of easing the interpretation of results.}
    \end{itemize}
  }
}
\\\\\\
\textbf{Paper organization}. In Section \ref{background} we report the background of this research. In Section \ref{research} we outline the research method followed along this study. We analyze a toy-experiment to show the extent to which variation of quality scores may pass unnoticed in SE experiments in Section \ref{simulation}. We analyze a real-life industrial experiment on Test-Driven Development in Section \ref{experiment}. We discuss our findings in Section \ref{discussion}. We outline the threats to validity of this study in Section \ref{threats}. Finally, we show the conclusions of this study in Section \ref{conclusion}.

\section{Background}
\label{background}

SE experiments' results are usually conveyed in terms of effect sizes, $p$-values and confidence intervals (CIs) \cite{cumming2013understanding, field2013discovering}. Effect sizes quantify the relationship between two groups (or more generally, between two variables: the dependent and the independent variable \cite{borenstein2011introduction}). Effect sizes can be provided in either \textit{standardized units} (e.g., Cohen's d that conveys the difference between the means of two groups divided by a pooled standard deviation  \cite{borenstein2011introduction}) or in \textit{unstandardized units} (e.g., $t$-tests' estimates that convey the difference between the means of two groups in natural units \cite{borenstein2011introduction}). $p$-values quantify the probability of achieving such effect size---or a larger one---given that a certain \textit{null hypothesis} (generally stating that there is no relationship between the dependent and independent variable) is true \cite{cumming2013understanding}. If the $p$-value is lower than a certain threshold (typically lower than 0.05 \cite{Cohen94theearth}), then it is claimed that the effect size is \textit{statistically significant}. If the effect size is statistically significant, then the relationship between the dependent and independent variable can be claimed to be different from 0---at least given the evidence collected from the experiment's data. Confidence intervals include the range of effect sizes compatible with the data at a certain probability threshold \cite{Kruschke2017}. Confidence intervals (CIs) are commonly used as a measure of \textit{precision} of the effect size \cite{cumming2013understanding}: the narrower the confidence interval at a certain threshold (e.g., 95\%), the larger the accuracy of the effect size and viceversa.

Traditionally used statistical tests---as well as traditionally used effect sizes such as Cohen's d \cite{fritz2012effect}---depend upon certain \textit{statistical assumptions} to provide reliable results \cite{field2013discovering}. For example, they depend upon the normality assumption and more critically, on the \textit{homogeneity of variances} assumption---when the groups are \textit{independent} \cite{field2013discovering}.\footnote{Even though other statistical tests allowing for unequal variances across groups are also available (e.g., the Welch's $t$-test, Generalized Least Squares, etc. \cite{bates2014fitting}), they are rarely used to analyze SE experiments \cite{dybaa2006systematic}, and thus, left out of our study.} In view of this, SE researchers routinely apply statistical tests (such as the Levene test, etc. \cite{field2013discovering}) to check the homogeneity of variances assumption and thus, being able to interpret the results of their experiments.

Unfortunately, obtaining a non-significant $p$-value by means of a statistical test such as the Levene test, and thus claiming that the data are compatible with the homogeneity of variances assumption, \textit{does not} imply that the effects of the technologies in the \textit{population} are equally sparse \cite{quinn2002experimental}. In particular, such misleading result may have just emerged as a consequence of the low statistical power of statistical tests in small sample sizes---as those common in SE experiments \cite{dybaa2006systematic}---and thus, the failure of the statistical test to detect differently sparse data across groups. 

Among the many statistical tests that can be used due to check differences across groups' variances (e.g., Barlett's, Hartley's, Levene's test, etc. \cite{quinn2002experimental}), along this study we illustrate the Brown-Forsythe test \cite{sheskin2003handbook} due to its robustness to departures from normality and its intuitiveness: it is just an ANOVA test performed on the deviations of each data point to the median of its group. The Brown-Forsythe test checks the \textit{null hypothesis} that the variances of all groups are identical. If the Brown-Forsythe test is statistically significant, then, there is enough evidence to claim that \textit{at least one} of the groups has a different variance \cite{sheskin2003handbook}. Not just statistical tests shall be run to identify differences across groups' variances: effect sizes quantifying the differences between them shall be also provided with the aim of easing the interpretation of results \cite{nakagawa2015meta}.  

Various effect sizes such as the \textit{lnCVR} are starting to be used to assess treatments' variances in ecology and medicine \cite{senior2016meta, stevens2016blood, nakagawa2015meta, senior2017dietary}. The lnCVR is more suitable than the difference between standard deviations for evaluating variances, specially in small sample sizes (as the sampling distribution of the standard deviation may not follow normality \cite{nakagawa2015meta}). The lnCVR stands for the \textit{natural logarithm (ln)} of the \textit{ratio between the coefficients of variation (CVR)} of two groups. The coefficient of variation (CV) of each group can be simply obtained by dividing its standard deviation by its mean (i.e., $sd/mean$). Large coefficients of variation (CVs) indicate large variance over small mean effects. On the contrary, small CVs indicate small variance over large mean effects. In SE parlance, a technology has a large CV if all developers perform wildly different to each other and perform to the minimum. On the contrary, a technology has a small CV if all developers perform similarly to each other and perform to the maximum. Thus, when obtaining the ratio between two CVs (i.e., obtaining the CVR), if the technology in the numerator is \textit{less variable} than that in the denominator, the natural logarithm of the CVR (i.e., the lnCVR) tends to a negative number. The larger the \textit{magnitude} of the lnCVR, the larger the difference between the variances of both technologies. 

For example, if we run an experiment to compare the performance of two technologies with regard to their variances (let us say Method A and Method B) and the CV of Method A is equal to 1, and the CV of Method B is equal to 2, this implies that Method A scores are \textit{half as variable} as those of Method B. When dividing their CVs and taking the natural logarithm (i.e., calculating the lnCVR), a negative number is obtained (i.e., ln(1/2)=-0.30). The larger the magnitude of the lnCVR (i.e., the more negative the number is), the larger the difference between the variances of both technologies. The lnCVR effect size is defined as follows \cite{nakagawa2015meta}:

$$ lnCVR= ln\left(\frac{CV_t}{CV_c}\right)  + \frac{1}{2*(n_t-1)} + \frac{1}{2*(n_c-1)} $$ 

where $CV_t$ and $CV_c$ are $s_t/\bar{x}_t$ and $s_c/\bar{x}_c$, for the treatment and control group, respectively.

\section{Research Method}
\label{research}

We conducted a literature review on the role of variance in ecology and medicine after realizing that variance is starting to be evaluated to assess the performance of new treatments in such disciplines \cite{senior2016meta, stevens2016blood, nakagawa2015meta, senior2017dietary}. In addition, from our experience at conducting and analyzing SE experiments, and after looking at SE experiments' reports included in Dyba et. al \cite{dybaa2006systematic}, we noticed that variances are rarely assessed in SE but to be able to interpret traditional statistical tests' results (e.g., the results of $t$-tests or ANOVAs \cite{dybaa2006systematic}).

With the aim of motivating the relevance of variance on technology performance, and illustrating visually why SE experiments' small sample sizes may not be able to detect differences across them, along this article we rely on simulation. In particular, we first simulate the performance of two hypothetical technologies in a continuous outcome (lets us say quality in a percentage scale) in an \textit{imaginary population} of developers (e.g., the population of all Finnish developers). We simulate the performance of each technology by means of a Beta distribution \cite{gelman2014bayesian}. Even though the shape of these distributions in the \textit{population} may never be known (unless, eventually after an infinite number of experiments, all the developers within the population had been sampled), by means of simulation \textit{we can act as if we knew the real distribution} of quality scores of each technology in the population, and then, simulate experiments just by sampling from these distributions (i.e., obtaining random data-points, each representing a different developer).

Beta distributions are a family of continuous probability distributions defined in the interval [0,1] with a shape governed by two parameters: $\alpha$ and $\beta$ \cite{gelman2014bayesian}. Both $\alpha$ and $\beta$ parameters define the shape (and thus, the mean and variance) of beta distributions \cite{gelman2014bayesian}. The relationship between means, variances, $\alpha$ and $\beta$ parameters follows:

$$ Mean= \frac{\alpha}{\alpha + \beta}$$

and 

$$ Variance= \frac{\alpha*\beta}{(\alpha+\beta)^2*(\alpha+\beta+1)} $$

\begin{figure}[h!]
\includegraphics[width=12cm,keepaspectratio]{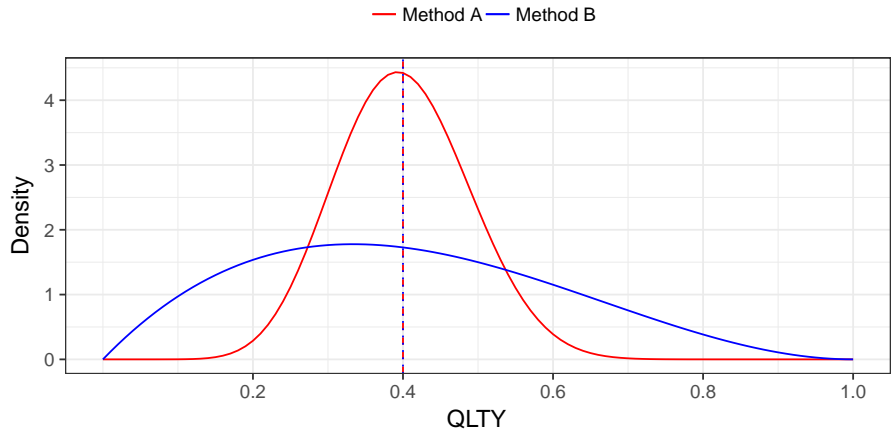}
\caption{Beta distributions.} \label{fig1}
\end{figure}

Thus, obtaining Beta distributions with identical means and different variances (or standard deviations) is straightforward by selecting appropriate $\alpha$ and $\beta$ parameters. As an example, Figure \ref{fig1} shows the \textit{Beta(12,18)} distribution (with $M=0.4;~SD=0.09$) and the \textit{Beta(2,3)} distribution (with $M=0.4;~SD=0.2$). For illustrative purposes, let us suppose that the \textit{Beta(12,18)} distribution plays the role of the quality scores achieved with Method A in \textit{a certain population} of developers, while the \textit{Beta(2,3)} distribution plays the role of the quality scores achieved with Method B in \textit{that same population}.

As it can be seen in Figure \ref{fig1}, Method A quality scores are clumped together around 0.4 (i.e., 40\%). In SE terms, most developers using Method A obtain quality scores around 40\%. On the other hand, developers using Method B obtain sparse quality scores (ranging between 0 and 100\%). Thus, according to the simulation parameters, both means are identical (see the overlapping dashed red and blue lines in $M=0.4$), even though the spread of the scores in Method B double those of Method A. Even though the means of Method A and B are identical, can we assume that both methods would perform similarly in a software project? Should a manager just look at means when deciding which technology to adopt in his company?

With the aim of portraying the insights that may be obtained in a prototypical SE experiment, we sample 15 data-points from each distribution (each data-point representing a different imaginary developer). This way, we obtain a \textit{simulated AB between-subjects experiment} \cite{wohlin2012experimentation} with a sample size of 30 (a common sample size in SE experiments according to Dyba et. al \cite{dybaa2006systematic}). Then, we analyze the experiment with the statistical test usually performed in such circumstances in SE  \cite{wohlin2012experimentation, juristo2013basics}: an independent $t$-test. The independent $t$-test relies on the normality and the homogeneity of variances assumptions \cite{field2013discovering}. We assess the normality assumption by means of the Shapiro-Wilk test. Then we check the homogeneity of variances assumption by means of the Brown-Forsythe test \cite{brown1974robust}. With the aim of easing the interpretation of results we calculate the Hedge's g effect size and the lnCVR effect size \cite{nakagawa2015meta}.

Finally, we analyze the results of a \textit{real-life industrial experiment} evaluating the performance of TDD on quality. We follow an identical procedure to that followed for analyzing the simulated experiment, but this time, we analyze the data with a dependent $t$-test instead (as the experiment uses an AB within-subjects design instead of an AB between-subjects design---see below). The dependent $t$-test relies on the normality assumption \cite{field2013discovering}. As in the case of the independent $t$-test, we assess the normality assumption by means of the Shapiro-Wilk test. Instead of stopping there and interpreting results as usual, \textit{we go a step further} and assess the differences between the variances by means of the Brown-Forsythe test. Finally, we complement the statistical analyses with their respective effect sizes. Did this last step reveal any extra insight on the performance of TDD? 

\section{Simulated Experiment}
\label{simulation}

\begin{figure}[h!]
\includegraphics[width=12cm,height=6cm]{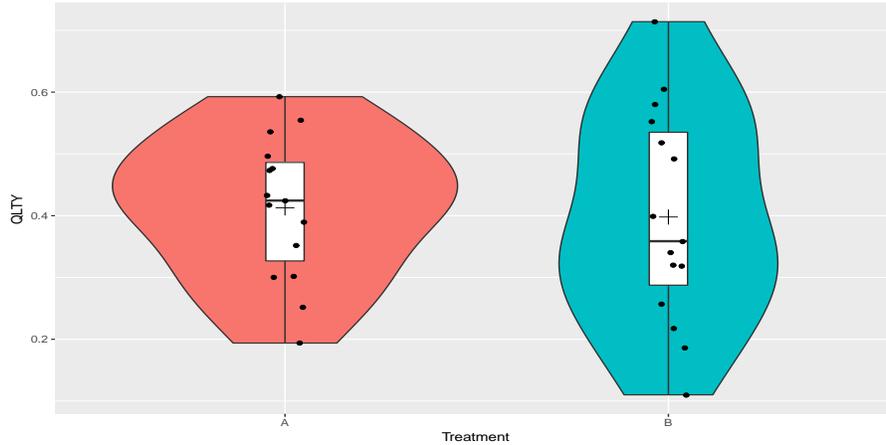}
\caption{Method A vs. Method B: violin-plot and box-plot.} \label{fig2}
\end{figure}

Figure \ref{fig2} shows the violin-plot and box-plot corresponding to a simulated AB between-subjects experiment comparing the performance of Method A and B. The data of each group have been simply obtained by sampling 15 different points from each of the two Beta distributions previously presented in Figure \ref{fig1} (Section \ref{research}). Table \ref{descriptive_statistics_simulation} shows the descriptive statistics of quality in each group.

As it can be seen in Figure \ref{fig2}, and as expected, most subjects applying Method A obtained quality scores clumped around 0.4, while subjects applying Method B obtained more sparse quality scores. By looking at Table \ref{descriptive_statistics_simulation}, it can also be seen that the ratio between the means of both groups is almost 1:1 ($M=0.413$ divided by $M=0.398$). On the contrary, the ratio between the standard deviations of both groups seems much larger (i.e., a ratio of 1:1.5 where Method B's standard deviation is larger than that of Method A). 

\begin{table}[h!]
\caption{Descriptive statistics.}
\label{descriptive_statistics_simulation}
\begin{center}
\begin{tabular}{lcccc} \hline \hline
\textbf{Treatment} & \textbf{N} & \textbf{Mean} & \textbf{SD} & \textbf{Median} \\ \hline
A & $15$ & $0.413$ & $0.115$ & $0.424$ \\ 
B & $15$ & $0.398$ & $0.173$ & $0.359$ \\ \hline
\end{tabular}
\end{center}
\end{table}

We run an independent $t$-test to analyze the data. The independent $t$-test requires the data to meet the normality assumption and the homogeneity of variances assumption. We used the Shapiro-Wilk test to check the normality assumption. According to the Shapiro-Wilk test, both distributions can be assumed to be normally distributed ($p$-value=0.92 and $p$-value=0.92, for Method A and B, respectively). We used the Brown-Forsythe's test to check the homogeneity of variance assumption. According to the Brown-Forsythe's test's results, the homogeneity of variance assumption is met, and then, both distributions can be assumed to be similarly sparse ($p$-value=0.147). Thus, despite having introduced different variances in the population by means of simulation, \textit{the Brown-Forsythe's test was unable to detect the difference between variances due to the small sample size} of the experiment. Thus, as the Brown-Forsythe's test says that the homogeneity assumption is met, the Shapiro-Wilk test says that the normality assumption is met, and if this was a real experiment we would have no idea about the shape of the distributions in the population, we proceed as usual and interpret the results of our experiment according to the results of an independent $t$-test. Table \ref{results_t_test} shows the results of the independent $t$-test.

\begin{table}[h!] \centering 
  \caption{Independent $t$-test for quality: Method A vs Method B.} 
  \label{results_t_test} 
\begin{tabular}{lccc} \hline \hline 
\textbf{Coeff.} & \textbf{Estimate} & \textbf{$t$-statistic} & \textbf{$p$-value} \\ \hline
Diff & -0.015 & -0.276 & 0.784 \\  \hline
\end{tabular} 
\end{table} 

As it can be seen in Table \ref{results_t_test}, the difference in performance between Method A and B is small ($M=-0.015$) and non-statistically significant ($p$-value=0.784). In addition, \textit{Hedge's g} is equal to $M=-0.0982$ (i.e. a small effect size according to rules of thumb \cite{borenstein2011introduction}). Thus, according to the results of the $t$-test---and the Hedge's g magnitude---\textit{the difference between the means of both methods in the population} (i.e., the dashed red and blue lines in Figure \ref{fig1}) \textit{is almost negligible}---as it was expected according to the parameters of the simulation. In SE parlance, \textit{Method A and Method B seem to perform similarly}.

Finally, the \textit{lnCVR} is equal to $M=0.445$. Back-transforming the lnCVR to natural scale (i.e., $exp(lnCVR)=1.504$), Method B's scores seem to be around 50\% (i.e., 1.504-1) more variable than those of Method A -as it was expected according to the parameters of the simulation. Thus, \textit{the scores of each method in the population seem differently spread according to the lnCVR}. However, and despite the large difference between the variances of both groups, \textit{the Brown-Forsythe's test was unable to detect the real difference between both methods' variances} (see above) due to the small sample size of the experiment.

\section{Real Experiment}
\label{experiment}

We run an experiment at a telecommunications company in 2014 with the aim of assessing the performance of TDD on external quality. The independent variable within the experiment is \textbf{development approach}, with TDD and ITL---the reverse-order methodology of TDD following Erdogmus et. al \cite{erdogmus2005effectiveness}---as treatments. We measured external quality as the percentage of test cases that successfully passed from a battery of test cases that we built to test participants' solutions. Specifically, QLTY was measured as: 
$$ QLTY =\frac{\#Test~Cases(Pass)}{\#Test~Cases(All)}*100 $$

\subsection{Experimental Settings}

Table \ref{tab:settings_baseline_replications} summarizes the settings of the experiment.

\begin{table}[h!]
\small
\begin{center}
\caption{Experimental settings.}
\label{tab:settings_baseline_replications}
\begin{tabular}{ p{3.8cm}  p{4.2cm} } \hline \hline
\textbf{Aspect} & \textbf{Values}  \\ \hline
\textbf{Factors} & Development Approach   \\ 
\textbf{Treatments} & TDD vs ITL \\ 
\textbf{Response variables} & QLTY\\ 
\textbf{Design} & AB Within-subjects\\ 
\textbf{Training} & TDD seminar\\ 
\textbf{Training duration} & 3 days/6 hours\\ 
\textbf{Experiment Duration} & 2.25 hours\\ 
\textbf{Environment} & C++, Eclipse, Boost Testing\\
\textbf{Number of subjects} & 20 \\ \hline
\end{tabular}
\end{center}
\end{table}

\subsection{Data Analysis}

\begin{figure}[h!]
\includegraphics[width=12cm,keepaspectratio]{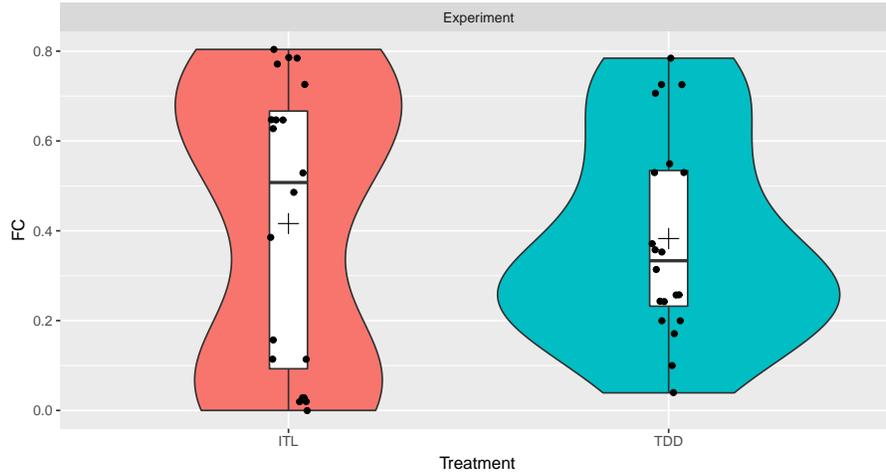}
\caption{ITL vs. TDD: violin-plot and box-plot.} \label{fig3}
\end{figure}

Figure \ref{fig3} shows the violin-plot and box-plot of the data gathered. Table \ref{descriptive_statistics_company} provides the corresponding descriptive statistics.

As it can be seen in Figure \ref{fig3}, ITL's quality scores look clumped either at the top or at the bottom of the distribution. In addition, TDD's quality scores seem grouped around the mean of the distribution (i.e., around 40\%).

\begin{table}[h!]
\caption{Descriptive statistics.}
\label{descriptive_statistics_company}
\begin{center}
\begin{tabular}{lcccc} \hline \hline
\textbf{Treatment} & \textbf{N} & \textbf{Mean} & \textbf{SD} & \textbf{Median} \\ \hline
ITL & $20$ & $0.416$ & $0.317$ & $0.508$ \\ 
TDD & $20$ & $0.383$ & $0.225$ & $0.333$ \\ \hline
\end{tabular}
\end{center}
\end{table}

As it can be seen in Table \ref{descriptive_statistics_company}, the ratio of the means is around 1:1 (i.e., ITL and TDD means are similar). However, the ratio of standard deviations is close to 1:1.4. In particular, TDD's quality scores seem less spread than those of ITL.

As usual in SE, we conduct a dependent $t$-test to analyze the data. The dependent $t$-test requires the difference of the quality scores between both groups to be normally distributed \cite{field2013discovering}. We check the normality assumption by means of the Shapiro-Wilk test. According to the Shapiro-Wilk test the difference between the quality scores of both groups is normally distributed ($p$-value=0.133). In view of this, we can interpret the results of the dependent $t$-test safely. Table \ref{dependent_t_test_result} shows the results of the dependent $t$-test.

\begin{table}[h!] \centering 
  \caption{Dependent $t$-test for QLTY: ITL vs. TDD.} 
  \label{dependent_t_test_result} 
\begin{tabular}{lccc} \hline \hline 
\textbf{Coeff.} & \textbf{Estimate} & \textbf{$t$-statistic} & \textbf{$p$-value} \\ \hline
Diff & -0.033 & -0.387 &0.703 \\  \hline
\end{tabular} 
\end{table} 

As it can be seen in Table \ref{dependent_t_test_result}, the difference in performance between ITL and TDD is small ($M=-0.033$) and non-statistically significant ($p$-value=0.703). In addition, the \textit{Hedge's g} is equal to $M=-0.119$ (i.e., a small effect size according to rules of thumb \cite{borenstein2011introduction}). In view of these results, \textit{TDD does not offer any advantage over ITL on quality}. 

However, in this occasion we go a step further than usual when analyzing SE experiments: instead of just interpreting the results of the dependent $t$-test and finalizing the data analysis, we go ahead and study the differences between the variances of ITL and TDD by means of the Brown-Forsythe's test. According to the Brown-Forsythe's test, the difference between the variances is statistically significant ($p$-value=0.04). In addition, the \textit{lnCVR} is equal to $M=-0.260$. Back-transforming the lnCVR to natural scale (i.e., $exp(lnCVR)=0.77$), TDD scores are 33\% (1-0.77) less variable than those of ITL. Thus, \textit{subjects with TDD seem to achieve more consistent quality scores than with ITL}. In view of this, \textit{TDD does offer advantages over ITL} (see below).

\section{Discussion}
\label{discussion}

As we have seen in the simulated experiment, meeting the homogeneity of variances assumption according to a statistical test (e.g., Levene,  Brown-Forsythe, etc.) \textit{does not} imply that technologies have identical variances: the presence of small sample sizes---as it is common in SE experiments \cite{dybaa2006systematic}---makes statistical tests under-powered to detect real differences \cite{quinn2002experimental}. Besides, as we have seen in the industrial experiment that we analyzed, not needing to check the homogeneity of variances assumption (as in the dependent $t$-test \cite{field2013discovering}) \textit{does not} imply that variances should be overlooked when judging the performance of new technologies: technologies with identical means may turn to be more---or less---beneficial depending upon their variances.

In the particular case of the industrial experiment that we analyzed, TDD seemed to provide less sparse quality scores than ITL---despite their similar means. If we had relied just on traditional statistical tests' results ---as it is commonly done in SE---not much could have been said: ITL seemed to perform similarly to TDD. However we went a step further: instead of solely relying on the results of the dependent $t$-test to judge the performance of TDD, we also analyzed the data to uncover differences across variances by means of the Brown-Forsythe's test. According to its results, TDD provided significantly less sparse quality scores than ITL. In SE terms, \textit{the quality scores achieved with TDD seemed less dependent upon developers' characteristics than those achieved with ITL} (as TDD quality scores resemble much more to each other than those of ITL regardless of the developers' characteristics). This has also some implications at the management level: if an hypothetical manager selects ITL for her group of developers, \textit{the quality scores of her developers can be hardly predicted} (as they may be either low or high quality scores). In turn, in such group of developers, some developers may achieve large quality scores (perhaps those assigned to develop some part of a software product) and some others may achieve small quality scores (perhaps those assigned to other part of a software product). Depending upon the assignment of developers to functionality, \textit{this may be detrimental to the construction of a new software product} (e.g., when the developers achieving the worse quality scores are assigned unknowingly to develop the core functionality of the software product). On the contrary, if the manager had selected TDD for her group of developers, the quality scores of all developers may have resembled more to each other and thus, the quality of the whole software product may have been more similar across its functionalities.

Notice that even though we were able to identify real differences across technologies variances in the industrial experiment that we conducted, this may not be the case in most SE experiments due to their typical small sample sizes \cite{dybaa2006systematic}. We could see this in the simulated experiment, where despite different variances were \textit{designed} within the population, it was not possible noticing the difference between them in the experiment. Thus, we suggest, as well as effect sizes such as Cohen's d are commonly provided for quantifying differences across treatment means, variance effect sizes such as lnCVR should also be provided to eventually ease the identification of real differences across treatment variances \cite{nakagawa2015meta}. 

The main message of this article is that variance can be a differentiating element when assessing the performance of new technologies. Technologies with large variances may imply unpredictable performances and "developer-dependent" characteristics (e.g., skills, background etc.) impacting results. On the contrary, technologies with small variances may imply predictable performances and "robustness" to developers' characteristics. In view of this, we suggest to analyze variance in SE experiments to uncover perhaps "hidden" strengths---or weaknesses---of the technologies under assessment.

Finally, we want to highlight than whenever an experiment is being analyzed by means of a traditional statistical test such as the $t$-test, what is being compared is not the performance of the two technologies in general, but instead, the difference \textit{between the means} of the two technologies. Put differently, sample-to-population inferences are being made on \textit{statistical points} (e.g., differences between means) and not on the distribution of the data. In view of this, not just differences between means may be of interest to judge technology's performance: differences between variances, medians or even quantiles may be also of interest. Even though this article was just a call to analyze variance in SE experiments, by using more advanced statistical methods such as Bootstrap \cite{quinn2002experimental}, it is also possible to assess differences between variances, medians, quantiles or even customized statistical points. Are we going to just rely on means to judge the performance of new technologies? After all, now we may be at the verge of assessing the performance of new technologies under perspectives never thought before.

\section{Threats to Validity}
\label{threats}

\textit{Just one experiment, how generalizable are our results?} We acknowledge the limitations of our study with regard to the use of a single experiment. However, we have complemented the results of the real-life experiment with those of a simulation. In particular, we used this simulation to illustrate that similar circumstances to that of the real experiment may materialize unknowingly, and that just relying on traditional statistical tests' results to judge technology performance may be misleading. Both the sample size of the simulation and the sample size of the industrial experiment are representative of SE experiments according to Dyba et. al \cite{dybaa2006systematic}. Under this point of view, we expect our results to be representative for SE experiments. 

\textit{Parametric tests and effect sizes, are there any threats to their application?} Along this study we just relied upon \textit{parametric} tests and effect sizes to analyze the data. Even though they may be unsuitable to analyze non-normal data---as that common in SE experiments \cite{kitchenham2017robust}---we relied on them as they allow to provide inferences in terms of differences between \textit{statistical parameters} (e.g., means, variances, etc.) and \textit{not in terms of ranks} (as non-parametric tests such as the Wilcoxon or U-Mann Whitney do) \cite{field2013discovering}, they are usually recommended to analyze SE experiments \cite{wohlin2012experimentation, juristo2013basics}, they have been the most used to analyze SE experiments \cite{dybaa2006systematic}, and they are robust to departures from normality---even in smaller data-sets than those typical in SE experiments \cite{vickers2005parametric, norman2010likert}.  

\textit{Just one statistical test per statistical assumption, how limited are the findings?} Along this study we just used the Shapiro-Wilk test to check the normality assumption and to the Brown-Forsythe test to check the homogeneity of variances assumption. Even though we could have also used other tests such as the Kolmogorov-Smirnov test to check the normality assumption \cite{field2013discovering}, we used the Shapiro-Wilk test just for illustrative purposes. Besides, we used the Brown-Forsythe test as it is more suitable than the Levene test when data departs from normality \cite{sheskin2003handbook} (and thus, may be more suitable to analyze SE experiments). As an aside, in the simulated experiment and the real-life experiment that we analyzed both tests for normality and for homogeneity of variances provided similar results. Thus, results seem consistent regardless of the statistical test used.

\section{Conclusion}
\label{conclusion}

Commonly applied statistical tests to analyze SE experiments (e.g., $t$-test, ANOVA etc. \cite{wohlin2012experimentation, juristo2013basics}) rely on the homogeneity of variances assumption to provide sample-to-population inferences \cite{field2013discovering}. As a consequence, different variances across groups in SE are not seen as an opportunity to provide insights on technology's performance but instead, as a hindrance towards the analysis of the data \cite{dieste2011comparative, kitchenham2017robust}. Perhaps inadvertently, and as a consequence of just relying on traditional tests results, SE technologies are only assessed with regard to their mean performance.

Along our study we showed that meeting traditional statistical assumptions (e.g., homogeneity of variances) does not imply that the underlying data distributions are equally sparse. Instead, this may be a sign that a larger sample size is required to find statistical significant differences across technologies' performances. In addition, not needing to meet the homogeneity of variances assumption does not imply that variance should be overlooked when judging technology performance. 

Instead of considering variance as a hindrance, we suggest, variance should be considered in SE experiments as a valuable source of knowledge ---as it is already being done in other disciplines such as medicine or biology \cite{senior2016meta, stevens2016blood, nakagawa2015meta}. In particular, technologies with similar means may not be equally desirable if variances are largely dissimilar, specially if it is aimed at lowering the risk of adopting them in real practice. Under this point of view, technologies that make subjects resemble to each other may be more desirable than those technologies that do not (as not much deviation from the 'average' performance is expected with their use). Are you going to continue judging the relevance of technologies just in terms of means? Or are you going to move beyond them? After all, now you are a few extra statistical tests and effect sizes away from obtaining new insights on technology's performance.

\section*{Acknowledgments}
This research was developed with the support of the Spanish Ministry of Science and Innovation project TIN2014-60490-P.

\bibliographystyle{splncs}
\bibliography{main}

\end{document}